# Equilibrium properties of assembly of interacting superparamagnetic nanoparticles

N. A. Usov[1,2], O. N. Serebryakova[2]

[1]National University of Science and Technology «MISiS», 119049, Moscow, Russia
[2]Pushkov Institute of Terrestrial Magnetism, Ionosphere and Radio Wave Propagation, Russian Academy of Sciences, IZMIRAN, 142190, Troitsk, Moscow, Russia

**Abstract**

The stochastic Landau-Lifshitz equation is used to investigate the relaxation process and equilibrium magnetization of interacting assembly of superparamagnetic nanoparticles uniformly distributed in a nonmagnetic matrix. For weakly interacting assembly the equilibrium magnetization is shown to deviate significantly from the Langevin law in the range of moderate and large magnetic fields due to the influence of magnetic anisotropy energy. For dense assemblies with noticeable influence of the magneto- dipole interaction a significant dependence of the initial susceptibility on the assembly density is revealed. The difference between the initial susceptibility and the corresponding Langevin susceptibility can serve as an indication of the influence of the magneto- dipole interaction on the assembly properties. A new self-consistent approach is developed to explain the effect of mutual magneto- dipole interaction on the behavior of dense assembly of superparamagnetic nanoparticles. The probability densities of the components of random magnetic field acting on magnetic nanoparticles are calculated at thermodynamic equilibrium. The self-consistent probability densities of these components are found to be close to Gaussian distribution. It is shown that a decrease in the equilibrium assembly magnetization as a function of density can be explained by the disorienting effect of the random magnetic field on the particle magnetic moments.

## Introduction

Assemblies of superparamagnetic nanoparticles are widely used in various fields of nanotechnology, in particular, in biomedicine, for magnetic resonance imaging, targeted drug delivery, purification of biological media from toxins, in magnetic hyperthermia, etc. [1-4]. However, the study of the physical properties of dense assemblies of magnetic nanoparticles is complicated by the influence of a strong magneto-dipole interaction between the nanoparticles [5–13]. Formally, the equilibrium properties of an assembly of superparamagnetic nanoparticles distributed in a rigid media can be studied on the basis of the Gibbs principle [14-18], if for a given temperature $T$ of a thermal bath and applied magnetic field $H_0$ the complete thermodynamic equilibrium is established for a finite time. For such assembly the equilibrium magnetization, $M_{eq} = M_{eq}(H_0,T)$, can be calculated as the derivative of the free energy with respect to the applied magnetic field [14-18]. For given $T$ and $H_0$ values the equilibrium magnetization and magnetic susceptibility of a dense assembly with an average particle diameter $D$ should depend both on the particle magnetic parameters $(M_s,K)$, and on the assembly density, $\eta = N_p V/V_{cl}$. Here $M_s$ is the saturation magnetization, $K$ is the uniaxial anisotropy constant, $V = \pi D^3/6$ is the nanoparticle volume, and $N_p$ is the number of nanoparticles in a cluster of volume $V_{cl}$. Unfortunately, the direct use of the Gibbs statistical integral for calculating the equilibrium properties of dense nanoparticle assembly is associated with great mathematical difficulties.

In the classical paper [19] Langevin used Gibbs principle to calculate the equilibrium magnetization of non-interacting assembly of freely rotating magnetic dipoles. The Langevin law for equilibrium assembly magnetization is given in standard textbooks [20,21]. It is often used in the analysis of the experimental data [22-29]. However, it is important to recall that in the simplest Langevin approximation [19] both the magnetic anisotropy energy and the energy of the magneto- dipole interaction of the nanoparticles are neglected. Meanwhile, the influence of particle magnetic anisotropy on assembly behavior can be approximately investigated analytically [30,31] or numerically [32] based on the Gibbs formula in the limit of weakly interacting nanoparticles, $\eta \to 0$. On the other hand, the evaluation of the Gibbs statistical integral in the general case of interacting assembly is a difficult problem, well known in the theories of non-ideal gas, dipole fluids, plasma, and other fields of classical and quantum physics [14-18,20]. A similar problem also exists for interacting assemblies of superparamagnetic nanoparticles.

In recent years a significant amount of research [25, 33-55] has been devoted to theoretical and experimental study of the influence of the magneto- dipole interaction on the properties of dense nanoparticle assemblies. In particular, the Monte Carlo simulations have been carried out [25,33–42, 46,50,53] for assemblies of interacting magnetic nanoparticles in the temperature range exceeding the blocking temperature. Various generalizations of the Langevin formula were proposed to take into account the influence of the magneto- dipole interaction, such as the interacting superparamagnetic model (ISM) [43–45,51], or different versions of the effective magnetic field [46–50,52]. For the same purpose the thermodynamic perturbation theory [54] and the decomposition of the Gibbs statistical integral by the Born-Mayer method [55] were employed. However, despite the abundance of the approaches used, the understanding of the influence of the magneto- dipole interaction on the equilibrium properties of dense assembly of superparamagnetic nanoparticles seems still incomplete.

It is worth noting that in contrast to the theoretical problem mentioned, the measurement of the equilibrium magnetization and magnetic susceptibility of interacting assembly of superparamagnetic nanoparticles is a routine experimental task that can be performed using standard equipment [22-29]. In an attempt to improve theoretical understanding in this work the equilibrium magnetization of an assembly of interacting superparamagnetic nanoparticles uniformly distributed in a rigid nonmagnetic matrix is calculated by solving the stochastic Landau – Lifshitz (LL) equation [56–60]. This approach is an alternative to the classical method of Gibbs assemblies. It enables one to simultaneously take into account the effect of various types of magnetic anisotropy, magneto- dipole interaction, and thermal fluctuations of the particle magnetic moments on the assembly behavior. Moreover, this method allows one to consider also kinetic processes, such as the relaxation process to the equilibrium assembly magnetization.

Calculations based on the stochastic LL equation were performed in this work for a dilute assembly of nanoparticle clusters with a finite filling density $\eta$. The random positions of the nanoparticles in the cluster are assumed to be fixed, the rotation of the nanoparticles as a whole is excluded. The easy anisotropy axes of the particles are randomly oriented. The saturation magnetization of the particles is taken to be $M_s = 350$ emu/cm$^3$, that is typical for iron oxide nanoparticles [3,6,10], the uniaxial magnetic anisotropy constant varied in the range $K = 6\times10^4 - 1.5\times10^5$ erg/cm$^3$. The numerical simulations are carried out at a room temperature, $T = 300°$ K. Therefore, the diameter of spherical nanoparticles is restricted to the range $D < 25$ nm, to ensure [61] that the blocking temperature $T_b$ of the largest nanoparticles is well below the room temperature. Numerical calculations of the equilibrium magnetization and static susceptibility of a dilute assembly of dense clusters consisting of $N_p = 60 - 100$ nanoparticles are carried out in a range of applied magnetic fields, $H_0 = 0 - 600$ Oe, the cluster filling density being $\eta = 0 - 0.3$. A significant dependence of the assembly equilibrium magnetization on the intensity of the magneto- dipole interaction inside the clusters has been revealed.

In addition, the statistical properties of random magnetic field acting on magnetic nanoparticles in a dense assembly of superparamagnetic nanoparticles have been studied. Following the Lorentz approach [20], in a large equilibrium assembly of nanoparticles, which can be characterized by an average magnetization $M_{eq}(H_0,T)$, one can construct near typical nanoparticle a sphere (Lorentz sphere) with radius $R_L$ much larger than the average distance $L_{av}$ between nanoparticles. Outside Lorentz sphere the assembly magnetization can be considered approximately homogeneous, $<M(r)> = M_{eq}$. Therefore, near the center of the Lorentz sphere the magnetic field created by particles located outside the Lorentz sphere is close to zero. Consequently, the random component of magnetic field acting on a typical nanoparticle is determined only by the surrounding nanoparticles located inside the Lorentz sphere. In the present work the magnetic moment of a typical nanoparticle, and hence the equilibrium magnetization $M_{eq}$ of the assembly, is calculated self-consistently depending on the total magnetic field acting on the particle. The latter is the sum of the external magnetic field and the random magnetic field created by the surrounding nanoparticles located within the Lorentz sphere.

It is shown that the variant of the self-consistent field approximation developed in this work qualitatively correctly describes numerical simulation data for the equilibrium assembly magnetization in the entire range of applied magnetic fields investigated.

## Results and Discussion
### Dilute nanoparticle assembly

As emphasized in the Introduction, there are two important contributions that lead to a difference in the reduced equilibrium magnetization of interacting assembly, $m_{eq} = M_{eq}(H_0,T)/M_s$, from the Langevin law [19-21]

$$\frac{\langle M \rangle}{M_s} = m_L(x); \qquad m_L(x) = \coth(x) - \frac{1}{x}, \qquad (1)$$

where $x = M_s V H_0/k_B T$ is the dimensionless Langevin variable, $k_B$ is the Boltzmann constant. These are magnetic anisotropy energy and the energy of the magneto- dipole interaction. Let us discuss the influence of these factors on the equilibrium properties of the assembly separately.

Consider first the relatively simple case of a dilute nanoparticle assembly, $\eta \to 0$, neglecting the influence of magneto- dipole interaction. In this case the equilibrium assembly magnetization can be determined evaluating the Gibbs statistical integral [30-32]. The corresponding calculations for randomly oriented monodispersive assemblies of superparamagnetic nanoparticles are shown in Fig. 1.

Fig. 1a shows the dependence of the equilibrium magnetization of a randomly oriented assembly on the average particle diameter. Fig. 1b gives the dependence of the static magnetic susceptibility of the assembly on the applied magnetic field. As can be seen from this figure, the static susceptibility decreases with increasing applied magnetic field and substantially depends on the average particle diameter. Moreover, as Fig. 1c shows, the second derivative of equilibrium magnetization with respect to the magnetic field shows a pronounced minimum in the region of low magnetic fields. The position of the minimum is a function of particle diameter.

It is useful to normalize applied magnetic field to the particle anisotropy field, $H_a = 2K/M_s$, that is, to introduce the reduced variable $h_e = H_0/H_a$. Then it can be shown [30-32] that the equilibrium magnetization of a dilute assembly is a universal function of $h_e$ that depends only on the reduced height of the particle potential energy barrier, $R_b = KV/k_BT$, so that $m_{eq0} = m_{eq0}(h_e, R_b)$. However, as Fig. 1d shows in the limit of small magnetic fields the equilibrium magnetization curve coincides with the Langevin function, Eq. (1), for all values of the anisotropy constant $K$. This is a consequence of the fact that in the limit $h_e \to 0$ the expansion

$$m_{eq0}(h_e, R_b) = \frac{2}{3} R_b h_e + ... = \frac{M_s V H}{3 k_B T} + ... \quad (2)$$

is valid [30,31]. As a result, the dependence of the equilibrium magnetization on the anisotropy constant $K$ in the region of small magnetic fields disappears. At the same time, according to Fig. 1d for moderate and large magnetic fields the difference between the equilibrium magnetization and the Langevin function is very significant. It follows from Eq (2) that the static magnetic susceptibility of the assembly in the low-field region does not depend on the magnetic anisotropy constant. Therefore, it is impossible to determine the value of the anisotropy constant $K$ by measuring the static susceptibility of a dilute assembly, $dm_{eq0}/dH_0$, in the limit $h_e \to 0$. At the same time, as we will see later, the static magnetic susceptibility of an interacting assembly differs significantly from the Langevin susceptibility. This important fact makes it possible to evaluate the effect of the magneto- dipole interaction on the equilibrium properties of an assembly.

The significant influence of the particle magnetic anisotropy energy on the behavior of dilute assembly of monodispersive nanoparticles in the range of moderate and high magnetic fields, and at temperatures not too high with respect to blocking temperature of the assembly $T_b$ was studied in detail both experimentally [23,25] and theoretically [30-32]. The area of parameters $H_0$ and $T$, where there is a noticeable deviation of the equilibrium assembly magnetization from the Langevin law was characterized [23,25] as anisotropic superparamagnetism. Unfortunately, in a number of recent experimental works (see, for example, Refs. 26–29), the experimental data for the equilibrium assembly magnetization are described by a weighted sum of Langevin functions. In this way the particle size distribution is taken into account, whereas the influence of the magnetic anisotropy energy is completely ignored.

**Assembly of dense 3D clusters**

As noted in the Introduction, the direct application of the Gibbs principle for calculating the equilibrium magnetization of an assembly of interacting nanoparticles is associated with significant mathematical difficulties. To overcome this difficulty, various theoretical methods were used [25,33–55]. The most convincing results were obtained by means of Monte Carlo simulations [33–42,46,50,53] for assemblies of interacting superparamagnetic nanoparticles uniformly distributed in a nonmagnetic media. However, a known drawback of this method is the difficulty in estimating the actual time for evolution of the assembly in a given magnetic field, as individual Monte Carlo steps do not correspond to real physical time [33]. As an alternative approach to the problem, in the given paper we use direct numerical simulation based on a solution of the stochastic LL equation [56-60]. This method traces the temporal dynamics of the particle magnetic moments simultaneously taking into account the effects of thermal fluctuations and the strong magneto- dipole interaction between the particles of the assembly. The details of numerical modeling of the kinetic properties of an assembly of magnetic nanoparticles using the stochastic LL equation are described below in the Methods section.

Fig. 2 shows the magnetization relaxation curves of randomly oriented assemblies of magnetic nanoparticles of various diameters in a given external magnetic field $H_0$ for various initial magnetization states. In the magnetization distribution designated as Z state, at time $t = 0$ the particles are magnetized along the applied magnetic field, whereas for the R initial state the magnetic moments of the nanoparticles are randomly oriented in space. Both initial distributions of the particle magnetic moments differ from thermodynamically equilibrium. The temporal evolution of the assembly magnetization for $t > 0$ is shown in Fig. 2. It is calculated by solving the stochastic LL equation with a sufficiently small numerical time step $\Delta t$ with respect to characteristic particle precession time $T_p$ [60].

To obtain the complete magnetization relaxation curve of an assembly to the equilibrium state a sufficiently large number of the numerical time steps must be taken. The thermodynamic equilibrium is considered to be achieved when the magnetic relaxation curve approaches a constant value, $m_{eq} = M_{eq}/M_s$, and fluctuates around this value with a small dispersion, as shown in the relaxation curves presented in Fig. 2. In the calculations performed the number of nanoparticles in the clusters is $N_p = 60 - 100$. To obtain statistically reliable results a large number of numerical experiments, $N_{exp} = 100 – 200$, is carried out with the same initial conditions. The average magnetization of a dilute assembly of clusters is calculated by averaging over the set of magnetic relaxation curves of individual clusters with independent realization.

Fig. 2a compares the magnetization relaxation curves from a uniformly magnetized state (Z state) in applied magnetic field $H_0 = 10$ Oe for a non-interacting and interacting assemblies of nanoparticles of the same diameter $D = 21$ nm. To obtain the statistically reliable results shown in Fig. 2a the numerical simulation data were averaged over $N_{exp} = 200$ independent numerical experiments. In every numerical experiment $N = 3 \times 10^6$ numerical steps were performed with a small time step $\Delta t = 1.26 \times 10^{-11}$ s, the phenomenological damping constant is assumed to be $\kappa = 0.5$.

In Fig. 2a the relaxation curve for a non-interacting assembly, $\eta = 0$, can be described by a time dependent exponent with a single relaxation time $\tau = 2 \times 10^{-5}$ s. This curve approaches a constant value, $m_{eq0}$ = 0.133, which coincides with the reduced equilibrium magnetization of the non-interacting assembly calculated using the Gibbs formula. At the same time, as Fig. 2a shows, the relaxation curve for an assembly of clusters with a filling density $\eta$ = 0.278 cannot be characterized by a single relaxation time. To approximate this curve at least two exponents with significantly different relaxation times should be used. Nevertheless, as Fig. 2a shows, at sufficiently large times this curve also approaches a constant value, $m_{eq}$ = 0.056. It is reasonable to take this value as the equilibrium magnetization of an assembly of clusters with a filling density $\eta$ = 0.278 in applied magnetic field $H_0$ = 10 Oe.

As Fig. 2a shows the presence of a magneto-dipole interaction leads to a decrease in the magnetization relaxation time at the fast initial stage, followed by a much slower stage of the full establishment of thermodynamic equilibrium, during which the average magnetization of the assembly already changes relatively weakly. The equilibrium magnetization in the assembly of clusters with a noticeable intensity of the magneto-dipole interaction always decreases compared to that of the corresponding assembly of non-interacting nanoparticles.

This conclusion is confirmed by the data presented in Figs. 2b, 2c were the magnetization relaxation curves of various assemblies are shown for different initial states, i.e. Z and R initial states, respectively. As can be seen from Figs. 2b, 2c, in accordance with the Gibbs principle the equilibrium state of the assembly in a given applied magnetic field turns out to be the same, regardless of the type of initial magnetization configuration. It is worth mentioning that the Gibbs postulate is not applicable to study the temporal evolution of the assembly magnetization. Fortunately, it can be done numerically by solving the stochastic LL equation. The equilibrium value of the reduced magnetization of the assembly can be obtained by averaging the relaxation curve over a finite interval of times exceeding the characteristic time of magnetic relaxation, $t > \tau$.

For an assembly with given parameters ($D$, $M_s$, $K$, $N_p$, $\eta$) it is possible to obtain the equilibrium value of the reduced magnetization as a function of applied magnetic field using the calculations similar to those shown in Fig. 2b, 2c. The results of these calculations are shown in Fig. 3. As Fig. 3 shows, with an increase in the cluster filling density $\eta$, i.e. with increase in the intensity of the magneto-dipole interaction inside the clusters, the value of the equilibrium assembly magnetization decreases. According to Figs. 3a - 3c the magnetic susceptibility of the assembly, $dm_{eq}/dH_0$, in the low-field region, $H_0 \to 0$, substantially decreases as a function of the cluster filling density $\eta$. It is worth mentioning that for given magnetic parameters $M_s$ and $K$ and a temperature $T$ = 300° K for assemblies of nanoparticles with diameters $D \leq 21$ nm the reduced equilibrium magnetization vanishes in the limit $H_0 \to 0$. These assemblies exhibit typical superparamagnetic behavior. At the same time, as Fig. 3d shows, for the assembly of nanoparticles of larger diameter, $D$ = 25 nm, there is a remanent magnetization in the limit $H_0 \to 0$. Therefore, the blocking temperature of this assembly exceeds the room temperature value $T$ = 300° K. As a result, the true equilibrium state for this assembly is not reached for the finite evolutionary time. It is also interesting to note that according to Fig. 3d the remanent magnetization of the assembly decreases with increasing intensity of the magneto-dipole interaction.

Fig. 4 demonstrates an interesting universal behavior of the equilibrium magnetization curves for assemblies of nanoparticles with a noticeable intensity of the magneto-dipole interaction, $\eta \gtrsim 0.2$. While the equilibrium magnetization curves of assemblies of non-interacting nanoparticles substantially depend on the average particle diameter $D$ (see Fig. 1a), for interacting assemblies these curves practically coincide at the same cluster filling density $\eta$. An exception is the magnetization curves for particles with a diameter of $D$ = 25 nm, for which the remanent magnetization is nonzero.

To explain this effect, one notes that to an order of magnitude the magnetic anisotropy energy of the particle is $W_a \sim KV$, whereas the characteristic energy of the magneto-dipole interaction of the particles can be estimated as $W_m \sim (M_s V)^2 / L_{av}^3$, where $L_{av}$ is the average distance between the particles of the cluster. The latter can be estimated from the relation $L_{av}^3 = V_{cl}/N_p$, so that for the characteristic energy of the magneto-dipole interaction one obtains $W_m \sim M_s^2 V \eta$. Therefore, the energy ratio $W_a/W_m \sim K/M_s^2 \eta$ is independent of the nanoparticle volume and is approximately constant for a fixed $\eta$ value.

Fig. 5a shows the reduced equilibrium magnetization of the assemblies of nanoparticles with the same diameter $D$ = 21 nm, but with different magnetic anisotropy constants. It is noteworthy that the equilibrium magnetization of interacting assemblies differs significantly from the Langevin curve. Moreover, as Fig. 5a shows for sufficiently dense assemblies with cluster filling density $\eta$ = 0.278 the influence of particle magnetic anisotropy on the equilibrium magnetization curve is not significant.

In particular, the static magnetic susceptibility of the assembly, $dm_{eq}/dH_0$, in the limit $H_0 \to 0$ is practically independent of the value of the magnetic anisotropy constant, similar to the case of assembly of non-interacting nanoparticles (see Fig. 1d). However, the static magnetic susceptibility of the interacting assembly is significantly less than the Langevin value, $dm_{eq}/dH_0 = M_s V/3k_B T$.

To demonstrate clearly the effect of the magneto-dipole interaction on the equilibrium properties of the assembly, it is of interest to study the equilibrium magnetization curves of an assembly of nanoparticles with a negligibly small magnetic anisotropy constant, $K$ = 0. As Figs. 5b, 5c show, the equilibrium magnetization of the assemblies with $K$ = 0 approaches the Langevin curve only in the limit $\eta \to 0$. It is interesting to note that the magnetic susceptibility of such an assembly in the limit $H_0 \to 0$ substantially depends on the cluster filling density $\eta$. As can be seen from Fig. 3, this conclusion is also valid for assemblies with a finite anisotropy constant $K$. Therefore, the difference in the static magnetic susceptibility of the assembly from the Langevin value $dm_{eq0}/dH_0 = M_sV/3k_BT$ [30,31] indicates the influence of the magneto- dipole interaction of nanoparticles on the assembly properties.

**Self consistent field approximation**

The detailed numerical calculations performed above make it possible to quantitatively assess the change in the equilibrium and kinetic properties of the assembly with an increase in the intensity of the magneto- dipole interaction. However, numerical calculations do not shed light on the physical cause of such changes. It is clear that in the presence of a magneto- dipole interaction the magnetic field acting on a typical nanoparticle differs from the magnetic field $H_0$ applied to the assembly, since the magnetic fields of the surrounding nanoparticles also act on this nanoparticle. In dense clusters, at small distances between the nanoparticles, the magnetic fields of the nearest nanoparticles can be very significant. Therefore, of fundamental interest is the determination of the probability density of a random magnetic field acting on a typical magnetic nanoparticle of the assembly.

In recent years various approaches were proposed [46-52] to introduce effective magnetic field acting on a typical nanoparticle in a dense superparamagnetic assembly. However, it was shown [53] that the expressions suggested for the effective magnetic field in some cases are hardly consistent with the Monte Carlo simulation results. In this paper, we develop another approach to evaluate the effect of random magnetic field acting in a dense nanoparticle assembly.

Consider a sufficiently large spherical assembly shown schematically in Fig. 6, and select around a typical nanoparticle a spherical region (Lorentz sphere) of sufficient radius, $R_L \gg L_{av}$. Outside Lorentz sphere one can introduce a nearly homogeneous magnetization distribution close to the average assembly magnetization. Then, inside the Lorentz sphere, at least near its center, the magnetic field of external magnetic dipoles is almost completely compensated and close to zero [20]. Therefore, the magnetic field in the center of the Lorentz sphere acting on the reference particle is created by the surrounding nanoparticles located inside the Lorentz sphere.

First, let us analyze the probability density of random magnetic field acting on a typical particle of an assembly with a negligibly small magnetic anisotropy constant, $K = 0$. As Figs. 5b, 5c show, in such an assembly due to the influence of the magneto- dipole interaction the difference between the equilibrium magnetization and the Langevin law can be very large. Let $\mathbf{H} = (H_x, H_y, H_z)$ be the vector of the random magnetic field in the center of Lorentz sphere created by the particles located inside it. Without loss of generality one can assume that the external magnetic field $H_0$ is applied along the $Z$ axis of the Cartesian coordinates. Then, the total magnetic field in the center of Lorentz sphere is given by $\mathbf{H}_t = (H_x, H_y, H_z + H_0)$. Let $H_t$ be the module of this vector. It is reasonable to assume that in thermodynamic equilibrium the time-average magnetic moment of the reference particle located in the center of Lorentz sphere is

$$\langle M \rangle / M_s = m_L\left(\frac{M_s V H_t}{k_B T}\right),$$
$$H_t = \sqrt{H_x^2 + H_y^2 + (H_z + H_0)^2}, \qquad (3)$$

where $m_L(x)$ is the Langevin function, Eq. (1). The direction of the particle average magnetic moment is parallel to vector $\mathbf{H}_t$, so that

$$\langle M_x \rangle / \langle M \rangle = H_x / H_t ;$$
$$\langle M_y \rangle / \langle M \rangle = H_y / H_t ;$$
$$\langle M_z \rangle / \langle M \rangle = (H_0 + H_z) / H_t. \qquad (4)$$

Further, let $P(H_x, H_y, H_z)$ be the probability density of a random magnetic field created by surrounding particles in the center of the Lorentz sphere. Then, the average magnetization of the assembly in the direction of the applied magnetic field is given by

$$\frac{\langle M_z \rangle}{M_s} = \iiint m_L\left(\frac{M_s V H_t}{k_B T}\right)\frac{H_z + H_0}{H_t} P(H_x, H_y, H_z) dH_x dH_y dH_z.$$
(5)

Thus, to calculate the equilibrium assembly magnetization in the given approximation it is necessary to determine the probability density of random magnetic field in the center of the Lorentz sphere, created by nanoparticles located inside this sphere.

For given assembly parameters, a self-consistent probability density of the random magnetic field $P(H_x, H_y, H_z)$ can be obtained numerically by conducting a sufficient number of numerical experiments with random spherical clusters of a fixed volume $V_{cl}$, the number of particles $N_p$, and with the same filling density $\eta$. As will be shown below, the partial empirical probability densities $P(H_x)$, $P(H_y)$, and $P(H_z)$ of the random functions $H_x$, $H_y$, and $H_z$ are close to the Gaussian distributions. It is reasonable to assume that due to the random nature of the magnetic field at the center of the Lorentz sphere, which is the sum of a large number of independent contributions of the magnetic fields of individual nanoparticles, there is a relation

$$P(H_x, H_y, H_z) = P(H_x)P(H_y)P(H_z). \qquad (6)$$

To find self-consistent partial probability densities $P(H_x)$, $P(H_y)$ and $P(H_z)$, an appropriate iterative procedure should be performed. At the first stage of this procedure we consider all particles inside the Lorentz sphere to be magnetized strictly parallel to the applied magnetic field, so that $<M_x> = 0$, $<M_y> = 0$, $<M_z> = M_s$. Under this assumption we obtain the empirical probability densities $P^{(1)}(H_x)$, $P^{(1)}(H_y)$ and $P^{(1)}(H_z)$ of the first approximation in the following manner. A sufficiently wide range of magnetic fields, $(-H_{max}, H_{max})$, is divided into a large number of intervals of the same length, $\Delta H \ll H_{max}$. Then a sufficient number of numerical experiments $N_{exp}$ are performed in which random clusters of $N_p$ nanoparticles are created independently. The total random magnetic field at the center of each cluster is calculated and the relative numbers of clusters for which the components of the random magnetic field $H_x$, $H_y$, and $H_z$ fall into each predefined interval $\Delta H$ are determined.

To obtain the partial probability densities of the second approximation, we generate clusters in the volume of the Lorentz sphere in which the particle centers are randomly distributed, but the magnetization directions of individual nanoparticles are assigned in accordance with the probability density $P^{(1)}(H_x,H_y,H_z) = P^{(1)}(H_x)P^{(1)}(H_y)P^{(1)}(H_z)$. Namely, the magnetic field $\mathbf{H} = (H_x,H_y,H_z)$ acting on a specific nanoparticle of the cluster is set randomly, in accordance with the probability density $P^{(1)}(H_x,H_y,H_z)$. After that, the average magnetization of this particle is determined by Eqs. (3), (4). In this way, we can assign the magnetization of all $N_p - 1$ nanoparticles of the cluster and calculate the total magnetic field acting on the test particle located in the center of the Lorentz sphere. If we repeat this procedure a sufficient number of times, we can determine the empirical probability density in the second approximation, $P^{(2)}(H_x,H_y,H_z)$. These iterations are repeated until successively obtained probability densities, $P^{(i)}(H_x,H_y,H_z)$, $i = 1, 2, \ldots$ converge to a certain limit. This limiting probability density should be used in Eq. (5) to obtain the equilibrium magnetization of the assembly for a given value of external magnetic field $H_0$.

Calculations show that to obtain the probability density $P(H_x,H_y,H_z)$ with accuracy of about one percent, it is enough to carry out only 3-4 iterations of the iterative procedure described above. As a result of the first iteration, we obtain the partial probability densities of the first approximation, $P^{(1)}(H_x)$, $P^{(1)}(H_y)$ and $P^{(1)}(H_z)$, which are very close to the Gaussian distribution, $P(H) = \exp(-H^2/2\sigma^2)/\sqrt{2\pi}\sigma$, with some empirical standard deviations, $\sigma_x^{(1)}$, $\sigma_y^{(1)}$ and $\sigma_z^{(1)}$. As a result of the iterative procedure, we obtain sequences of empirical standard deviations, $\sigma_x^{(i)}$, $\sigma_y^{(i)}$ and $\sigma_z^{(i)}$, $i = 1, 2, \ldots$ which quickly converge to some limiting values, $\sigma_x$, $\sigma_y$ and $\sigma_z$. Moreover, due to the axial symmetry of the problem an approximate equality $\sigma_x^{(i)} \approx \sigma_y^{(i)}$ is satisfied at each iteration step.

As an example of the calculations performed, Fig. 7a shows the evolution of the empirical probability densities $P^{(1)}(H_x)$ - $P^{(4)}(H_x)$ for the $H_x$ component of random magnetic field during 4 successive stages of the iterative procedure. To obtain empirical probability density, at each stage of the iterative procedure $N_{exp} = 10^5$ numerical experiments were carried out in which spherical clusters consisting of $N_p = 60$ nanoparticles of diameter $D = 21$ nm were created, the cluster filling density being $\eta = 0.278$. To construct the empirical probability densities, the interval of magnetic fields (- 600, 600 Oe) was subdivided into 120 intervals of 10 Oe length each. The particles centers inside the cluster volume were randomly distributed using the algorithm described in Ref. 12 (see also Methods section). The particle magnetizations were assigned by means of the procedure described above and using Eqs. (3), (4). As can be seen from Fig. 7a, the successively obtained empirical probability densities, $P^{(i)}(H_x)$, $i = 1 - 4$, can be described with reasonable accuracy by the Gaussian distribution. The empirical standard deviations quickly converge to a constant limiting value. The empirical probability densities for the $H_y$ and $H_z$ components of the random magnetic field are of the same form. As Fig. 7b shows, for small values of the applied magnetic field, the limiting empirical standard deviations $\sigma_x$ and $\sigma_z$ turn out to be close. But with an increase in $H_0$ they begin to differ, but always $\sigma_x < \sigma_z$. Moreover, $\sigma_x = \sigma_y$ for the transverse components of the random magnetic field due to the axial symmetry of the problem.

Fig. 8 shows the examples of the calculation of the equilibrium assembly magnetization in the given approximation for assemblies with zero magnetic anisotropy constant, $K = 0$. In Figs. 8a, 8b, solid lines represent the results of direct numerical calculation of the equilibrium assembly magnetization using the stochastic LL equation for nanoparticles with diameters $D = 17$ and 21 nm, respectively. The dots show the corresponding results for the same quantity calculated in the self-consistent approximation developed. The number of nanoparticles in the Lorentz sphere in the latter case is fixed at $N_p = 60$, only 4 cycles of the iteration procedure being carried out for every dot in Figs. 8a, 8b.

As can be seen from Fig. 8a, 8b the maximum difference between the results of both calculations does not exceed 15%. The remaining difference between the two values is due to the presence of the correlation effects. Obviously, the dynamics of the magnetic moments of closely located nanoparticles should be strongly correlated, but this fact is not taken into account in the approximation developed. Fig. 8c shows the equilibrium magnetizations of random assembly of nanoparticles with $D = 21$ nm calculated for different numbers of nanoparticles in the Lorentz sphere. As can be seen from Fig. 8c, an increase in the number of particles in the Lorentz sphere in excess of $N_p = 60$ does not lead to any noticeable change in the equilibrium magnetization curve of the assembly.

As Fig. 7b shows, the difference between the self-consistent standard deviations $\sigma_x$ and $\sigma_z$ is usually small in a wide range of the applied magnetic field. Assuming approximately that $\sigma_x \approx \sigma_z = \sigma$ and performing calculations in a spherical coordinate system with the polar axis parallel to the direction of the applied magnetic field, one can rewrite Eq. (5) as follows

$$\frac{\langle M_z \rangle}{M_s} = \sqrt{\frac{2}{\pi}} \int_0^\infty m_L\left(\frac{M_s V H}{k_B T}\right) \exp\left(-\frac{H^2 + H_0^2}{2\sigma^2}\right) \times \frac{H^2(\xi \cosh\xi - \sinh\xi)}{\sigma^3 \xi^2} dH,$$

where the variable $\xi = HH_0/\sigma^2$. In the limit $H_0 \to 0$ this integral is estimated to be

$$\frac{\langle M_z \rangle}{M_s} \approx \frac{2}{3}\sqrt{\frac{2}{\pi}} \frac{m_L\left(\sqrt{3}M_s V \sigma(0)/k_B T\right)}{\sigma(0)} H_0, \quad (7)$$

where $\sigma(0)$ is the standard deviation at zero applied magnetic field. For characteristic values of the standard deviation, $\sigma(0) \sim 100$ Oe, the Langevin function $m_L$ in Eq. (7) changes slowly. Therefore, as Eq. (7) shows, with an increase in the parameter $\sigma$ the initial magnetic susceptibility of the assembly decreases approximately as $1/\sigma$. It can be shown that Eq. (7) accurately describes the initial linear portion of the curves $M(H)$ shown in Figs. 8 if one uses in Eq. (7) the corresponding $\sigma(0)$ values obtained by means of numerical simulation. Obviously, the decrease in the equilibrium magnetization of the assembly as a function of its density is due to the disorienting effect of the random magnetic field. Actually, under the influence of the random magnetic field the magnetic moments of the nanoparticles on average deviate from the direction of the applied magnetic field.

Similar calculations of the equilibrium assembly magnetization in the self-consistent field approximation were also performed for random assemblies with a finite value of the magnetic anisotropy constant $K$. Instead of using Eqs. (3), (4) in this case one has to assign the magnetizations of the nanoparticles within the Lorentz sphere by means of the corresponding Gibbs principle taking into account the value of the magnetic anisotropy constant and the directions of the easy anisotropy axes of every nanoparticle according to the formulas given in Ref. 32. Figs. 9a, 9b show the equilibrium magnetization for an assembly of nanoparticles with $K = 10^5$ erg/cm$^3$, $M_s = 350$ emu/cm$^3$ for the case of particles with diameters $D = 17$ and $21$ nm, respectively. As can be seen from Fig. 9, the magnetic field dependences of the equilibrium magnetizations obtained by two different methods turn out to be sufficiently close in the entire range of the applied magnetic fields studied.

For completeness, Fig. 10a shows the results of calculation of the equilibrium magnetization for a dilute assemblies of elongated and oblate clusters of magnetic nanoparticles with aspect ratios $D_z/D = 2.0$ and $D_z/D = 0.5$, respectively, in comparison with the results for a spherical cluster, $D_z/D = 1.0$. Here, $D_z$ and $D$ are the longitudinal and transverse diameters of the spheroidal magnetic cluster, respectively. For simplicity, external magnetic field is assumed to be applied along the axis of symmetry of the spheroidal clusters (Z axis). Fig. 10a shows the results of calculations of the equilibrium magnetization of a dilute assembly of clusters obtained by solving the stochastic LL equation. As can be seen from Fig. 10a, for a given value of the external magnetic field $H_0$, the equilibrium assembly magnetization increases for an elongated cluster with an aspect ratio $D_z/D > 1$ and decreases in the opposite case, $D_z/D < 1$.

It is known [62], that the demagnetizing field in a uniformly magnetized spheroid substantially depends on its demagnetizing factors, which are the functions of the spheroid aspect ratio $D_z/D$. For an elongated spheroid, $D_z/D > 1$, the longitudinal demagnetizing factor $N_z$ is less than the corresponding value for sphere, $N_{z0} = 4\pi/3$. Therefore, inside the Lorentz sphere created in an elongated spheroid, a uniform demagnetizing field acts, $H_d = (N_{z0} - N_z)M_s > 0$, which is added to the external uniform magnetic field $H_0$. In the case of an oblate spheroid, the demagnetizing field in the Lorentz sphere is directed against the external magnetic field, since for oblate spheroid the demagnetizing factor $N_z > N_{z0}$.

Thus, the results of calculations of the equilibrium magnetization shown in Fig. 10a are explained by the influence of the macroscopic demagnetizing field which acts inside the Lorentz sphere. In order to calculate the equilibrium magnetization of assembly of spheroidal clusters in the self-consistent field approximation, it is necessary to take into account the existence of a non-zero demagnetizing field inside the Lorentz sphere. For the case of elongated and oblate clusters with aspect ratios $D_z/D = 2.0$ and $D_z/D = 0.5$, respectively, the results of such calculations are shown in Fig. 10b.

**Conclusions**

An assembly of single-domain magnetic nanoparticles is a complex physical system whose properties are determined by many factors, such as the distribution of nanoparticles in size and shape, the density of the assembly, and the value of the main magnetic parameters of the particles. The behavior of the assembly depends also on the properties of the medium where the nanoparticles are distributed, as well as on the temperature and the magnitude of the applied magnetic field. It is worth noting that in contrast to classical plasma or quantum gases with Coulomb interaction [14–18] the anisotropic magneto- dipole interaction acts between the magnetic particles. Moreover, a superparamagnetic nanoparticle is characterized by an induced magnetization. The latter is close to zero in the absence of magnetic field acting on the particle, contrary to elementary particles whose electric charge is fixed.

In this paper we study the properties of a superparamagnetic assembly of monodispersive nanoparticles distributed in a solid nonmagnetic matrix. The calculations performed take into account magnetic anisotropy and the magneto- dipole interaction of particles, but the contact exchange interaction between the closest nanoparticles is ignored, supposing that the nanoparticles are protected by thin nonmagnetic shells. This model differs significantly from that describing ferrofluids [46,47,50,52,53]. In the latter case, one has to take into account the rotation of nanoparticle in a liquid as a whole, and also consider possible redistribution of nanoparticles in space with the formation of chains of particles and dense conglomerates [46,47, 50,52,53].

To realize the superparamagnetic regime the temperature of the medium should be higher than the characteristic blocking temperature $T_b$ of the particle magnetic moments. It is important that in a superparamagnetic assembly the relaxation to a thermodynamically equilibrium occurs for a finite observation time. The fundamental physical quantity of a superparamagnetic assembly is the equilibrium magnetization, $M_{eq} = M_{eq}(H_0, T)$, which can be easily measured experimentally [23-29]. Theoretically, this value can be determined on the basis of the Gibbs principle [14-21], as the derivative of the assembly's free energy with respect to the applied magnetic field. However, the direct calculation of the Gibbs statistical integral for an assembly of interacting magnetic nanoparticles is associated with great mathematical difficulties. In this paper, a new, physically adequate method is used for calculating the equilibrium magnetization of an assembly by solving the stochastic LL equation [56-60]. In contrast to the Monte Carlo calculations [25,33–42,46,50], the relaxation process to thermodynamic equilibrium in the assembly can be directly observed using the stochastic LL equation. Detailed calculations of the equilibrium magnetization were performed in this work for a dilute assemblies of magnetic clusters containing $N_p$ = 60 - 100 nanoparticles of the same diameter. The intensity of the magneto-dipole interaction inside the clusters can be controlled by changing the cluster filling density $\eta$.

For an assembly of weakly interacting nanoparticles it is shown that due to the influence of magnetic anisotropy energy, equilibrium magnetization differs significantly from the Langevin law in the range of moderate and large magnetic fields. This fact should be taken into account when analyzing experimental data for dilute assemblies. Nevertheless, for sufficiently small fields the dependence of the equilibrium magnetization on the magnetic anisotropy constant $K$ disappears. In this area the Langevin formula is valid and describes universal behavior of the assembly. For the assemblies of iron oxide nanoparticles studied in this paper the universal behavior is observed in the field range $H_0 \leq 50$ Oe. However, for dense assemblies with a noticeable influence of the magneto-dipole interaction a significant dependence of the initial susceptibility on the density is revealed. The difference between the initial susceptibility and the corresponding Langevin susceptibility can serve as an indication of the influence of the magneto-dipole interaction on the assembly properties.

In this paper a new approach to describe the influence of random magnetic field acting on particles in a dense assembly is proposed. In effective field theories [46-50,52] it is assumed that a typical nanoparticle of the assembly is subjected to some self-consistent magnetic field, which takes into account the influence of the magnetic fields of the surrounding nanoparticles. However, in a real assembly each nanoparticle is under the influence of its own local magnetic field which contains a random component. In this paper the probability densities of the components of random magnetic field acting on a typical magnetic nanoparticle are calculated. It is shown that self-consistent probability densities of these components are described by Gaussian distribution. Thus, the standard deviation in the Gaussian distribution becomes an important parameter of the theory. Knowing the probability density of the components of random magnetic field it is possible to calculate the equilibrium magnetization of the assembly in the given approximation as a function of applied magnetic field. It is shown that the approach developed satisfactorily describes the numerical results obtained for the equilibrium $M(H)$ curve with the help of stochastic LL equation.

The effect of intense magneto-dipole interaction on the properties of an assembly of magnetic nanoparticles is usually explained [25] either by a change in the characteristic height of energy barriers between potential wells of magnetic nanoparticles, or by some collective processes that simultaneously affect the magnetic state of closely spaced magnetic nanoparticles. Based on Eqs. (5), (6) in this work it is shown that a decrease in the equilibrium magnetization of an interacting nanoparticle assembly as a function of its density can be explained by the disorienting effect of random magnetic field. The latter, on average, leads to a deviation of the magnetic moments of the nanoparticles from the applied magnetic field direction. In this connection, it is worth noting that the broadening of spectral lines in a high temperature plasma was successfully explained by the action of a random electric microfield, the statistical properties of which are described by Holtsmark [63] or similar [64] distributions.

**Methods**
**Stochastic Landau- Lifshitz equation**

Dynamics of unit magnetization vector $\vec{\alpha}_i$ of $i$-th single-domain nanoparticle of the cluster is determined by the stochastic LL equation [56-60]

$$\frac{\partial \vec{\alpha}_i}{\partial t} = -\gamma_1 \vec{\alpha}_i \times (\vec{H}_{ef,i} + \vec{H}_{th,i}) - \kappa\gamma_1 \vec{\alpha}_i \times (\vec{\alpha}_i \times (\vec{H}_{ef,i} + \vec{H}_{th,i})), \quad (8)$$

where $\gamma$ is the gyromagnetic ratio, $\kappa$ is phenomenological damping constant, $\gamma_1 = \gamma/(1+\kappa^2)$, $\vec{H}_{ef,i}$ is the effective magnetic field and $\vec{H}_{th,i}$ is the thermal field. The effective magnetic field acting on a separate nanoparticle can be calculated as a derivative of the total cluster energy

$$\vec{H}_{ef,i} = -\frac{\partial W}{VM_s \partial \vec{\alpha}_i}. \quad (9)$$

The total magnetic energy of the cluster $W = W_a + W_Z + W_m$ is a sum of the magneto-crystalline anisotropy energy $W_a$, Zeeman energy $W_Z$ of the particles in applied magnetic field $\vec{H}_0$, and the energy of mutual magneto-dipole interaction of the particles $W_m$.

For spherical nanoparticles with uniaxial type of magnetic anisotropy the magneto-crystalline anisotropy energy is given by

$$W_a = K_1 V \sum_{i=1}^{N_p} \left[1 - (\vec{\alpha}_i \vec{e}_i)^2\right] \quad (10)$$

where $e_i$ is the orientation of the easy anisotropy axis of $i$-th particle of the cluster. Zeeman energy $W_Z$ of the cluster in applied magnetic field is given by

$$W_Z = -M_s V \sum_{i=1}^{N_p} \vec{\alpha}_i \vec{H}_0. \quad (11)$$

Next, for spherical uniformly magnetized nanoparticles the magnetostatic energy of the cluster can be represented as the energy of the point interacting dipoles located at the particle centers $r_i$ within the cluster. Then the magneto-dipole interacting energy is

$$W_m = \frac{M_s^2 V^2}{2} \sum_{i \neq j} \frac{\vec{\alpha}_i \vec{\alpha}_j - 3(\vec{\alpha}_i \vec{n}_{ij})(\vec{\alpha}_j \vec{n}_{ij})}{|\vec{r}_i - \vec{r}_j|^3}, \quad (12)$$

where $n_{ij}$ is the unit vector along the line connecting the centers of $i$-th and $j$-th particles, respectively.

Thus, the effective magnetic field acting on the $i$-th nanoparticle of the cluster is given by

$$\vec{H}_{ef,i} = H_a (\vec{\alpha}_i \vec{e}_i) \vec{e}_i + \vec{H}_0 - M_s V \sum_{j \neq i} \frac{\vec{\alpha}_j - 3(\vec{\alpha}_j \vec{n}_{ij}) \vec{n}_{ij}}{|\vec{r}_i - \vec{r}_j|^3}. \quad (13)$$

The thermal fields, $\vec{H}_{th,i}$, $i = 1,2, ..N_p$, acting on various nanoparticles of the cluster are statistically independent, with the following statistical properties [56] of their components for every nanoparticle

$$\begin{aligned}
\langle H_{th}^{(\alpha)}(t) \rangle &= 0; \\
\langle H_{th}^{(\alpha)}(t) H_{th}^{(\beta)}(t_1) \rangle &= \frac{2 k_B T \kappa}{\gamma M_s V} \delta_{\alpha\beta} \delta(t - t_1), \\
\alpha, \beta &= (x, y, z).
\end{aligned} \quad (14)$$

Here $\delta_{\alpha\beta}$ is the Kroneker symbol, and $\delta(t)$ is the delta function.

The procedure for solving stochastic differential equations (8), (13) and (14) is described in detail in Refs. 57 - 59.

**Random 3D clusters of nanoparticles**

It is worth noting that in Monte Carlo calculations performed to study the superparamagnetic nanoparticle assemblies the nanoparticle positions were randomly generated [36,40] on nodes of simple cubic lattices with a certain lattice parameter. This numerical algorithm can hardly be considered as truly random. In particular, it completely prevents the appearance of numerous and important assembly configurations where certain nanoparticles turn out to be very close to each other, i.e. closer than the lattice parameter chosen.

In the present study the 3D clusters consisting of $N_p$ identical magnetic particles with truly random positions were created using numerical algorithm developed in Ref. 12. First, a dense and approximately uniform set of $N$ random points $\{\rho_i\}$ was created in a sphere of the radius $R_{cl}$, so that $|\rho_i| \leq R_{cl}$, $i = 1,2, ...N$, for $N \gg N_p$. The first random point $\rho_1$ can be chosen as a center of the first nanoparticle of the assembly, $r_1 = \rho_1$. Then it is necessary to remove all points with coordinates $|\rho_i - r_1| \leq D$ from the initial set of the random points. Any random point in the remaining set of points can serve as a center of second nanoparticle of the assembly, for example, $r_2 = \rho_2$. Continuing this procedure, one can assign centers to all $N_p$ nanoparticles within the cluster volume. Moreover, not one of the nanoparticles of the assembly will be in direct contact with the surrounding particles. This algorithm allows one to create random 3D clusters of magnetic nanoparticles with filling densities $\eta < 0.5$. The orientations of the easy anisotropy axes $\{e_i\}$, $i = 1,2, ... N_p$, of nanoparticles in random 3D clusters were chosen randomly on the unit sphere.

**Acknowledgment**

The authors gratefully acknowledge the financial support of the Ministry of Higher Education and Science of the Russian Federation in the framework of Increase Competitiveness Program of NUST «MISIS», contract № K2-2019-012.

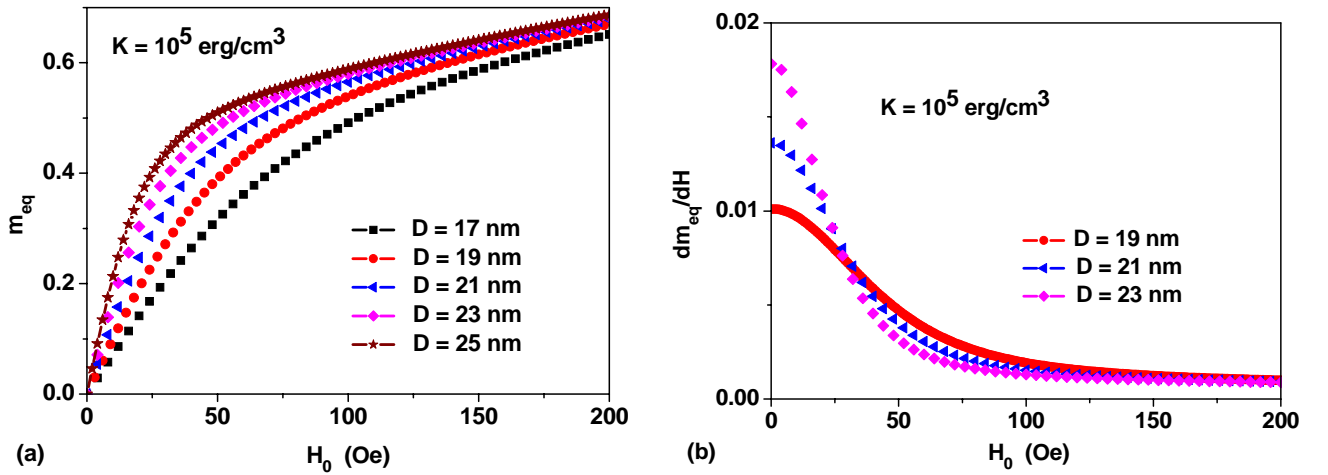

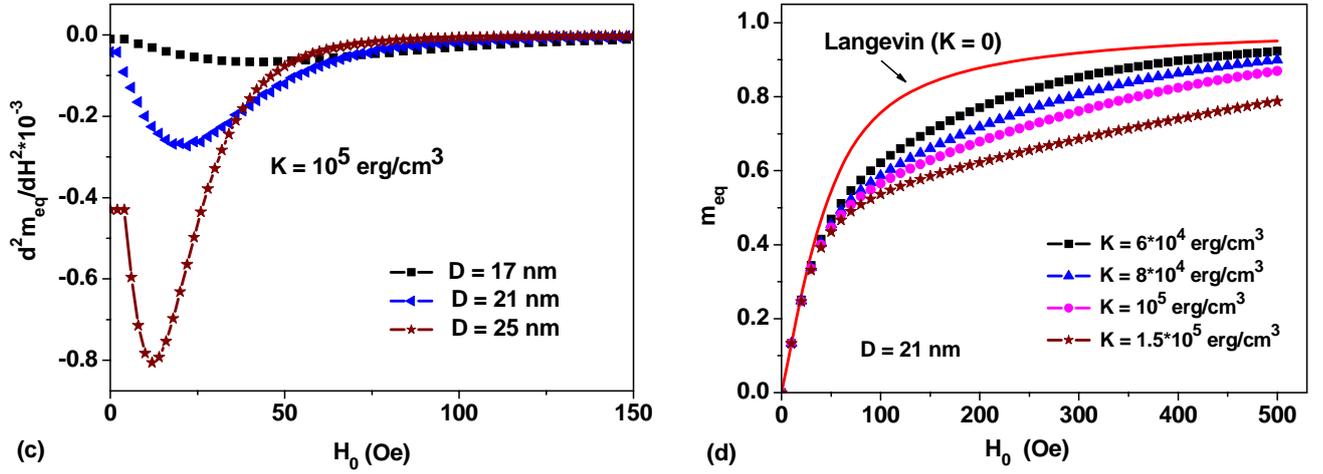

Fig. 1. a) The reduced equilibrium magnetization, $m_{eq} = M_{eq}/M_s$, of a randomly oriented assembly of non-interacting magnetic nanoparticles of different average diameters; b) reduced magnetic susceptibility of the assembly, $dm_{eq}/dH_0$; c) the second derivative of equilibrium magnetization, showing a pronounced minimum; d) the dependence of the reduced assembly magnetization on the value of the anisotropy constant $K$ for nanoparticles with a diameter $D = 21$ nm. Particle saturation magnetization $M_s = 350$ emu/cm$^3$, assembly temperature $T = 300$ K.

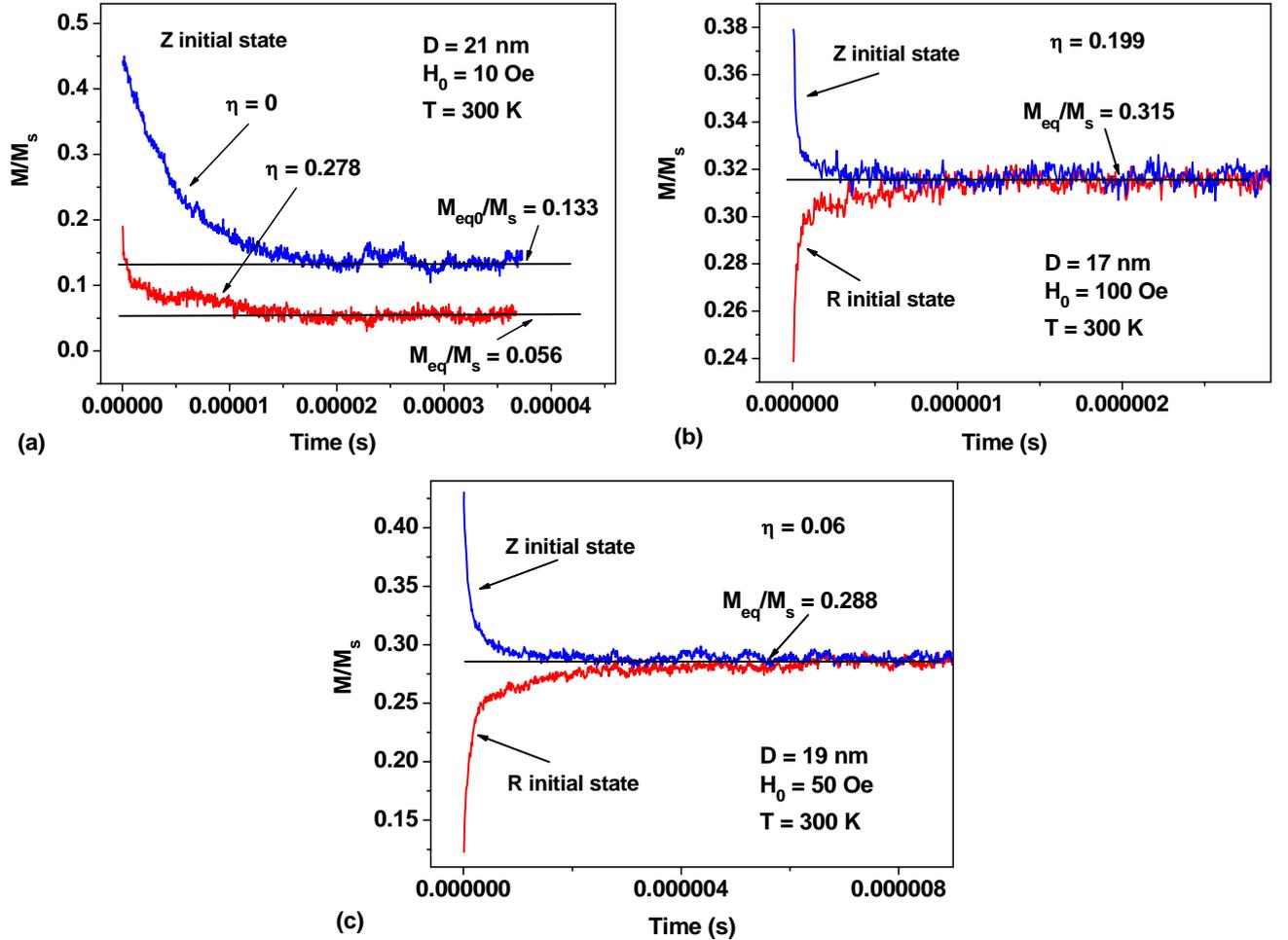

Fig. 2. Relaxation of magnetization to a thermodynamically equilibrium value in randomly oriented assemblies of magnetic nanoparticles: a) comparison of the magnetization relaxation curves of non-interacting ($\eta = 0$) and interacting ($\eta = 0.278$) assemblies of nanoparticles of diameter $D = 21$ nm; b), c) comparison of magnetization relaxation curves for different initial magnetization states for the assemblies of particles with diameters $D = 17$ and 19 nm, respectively. Particle saturation magnetization $M_s = 350$ emu/cm$^3$, magnetic anisotropy constant $K = 10^5$ erg/cm$^3$.

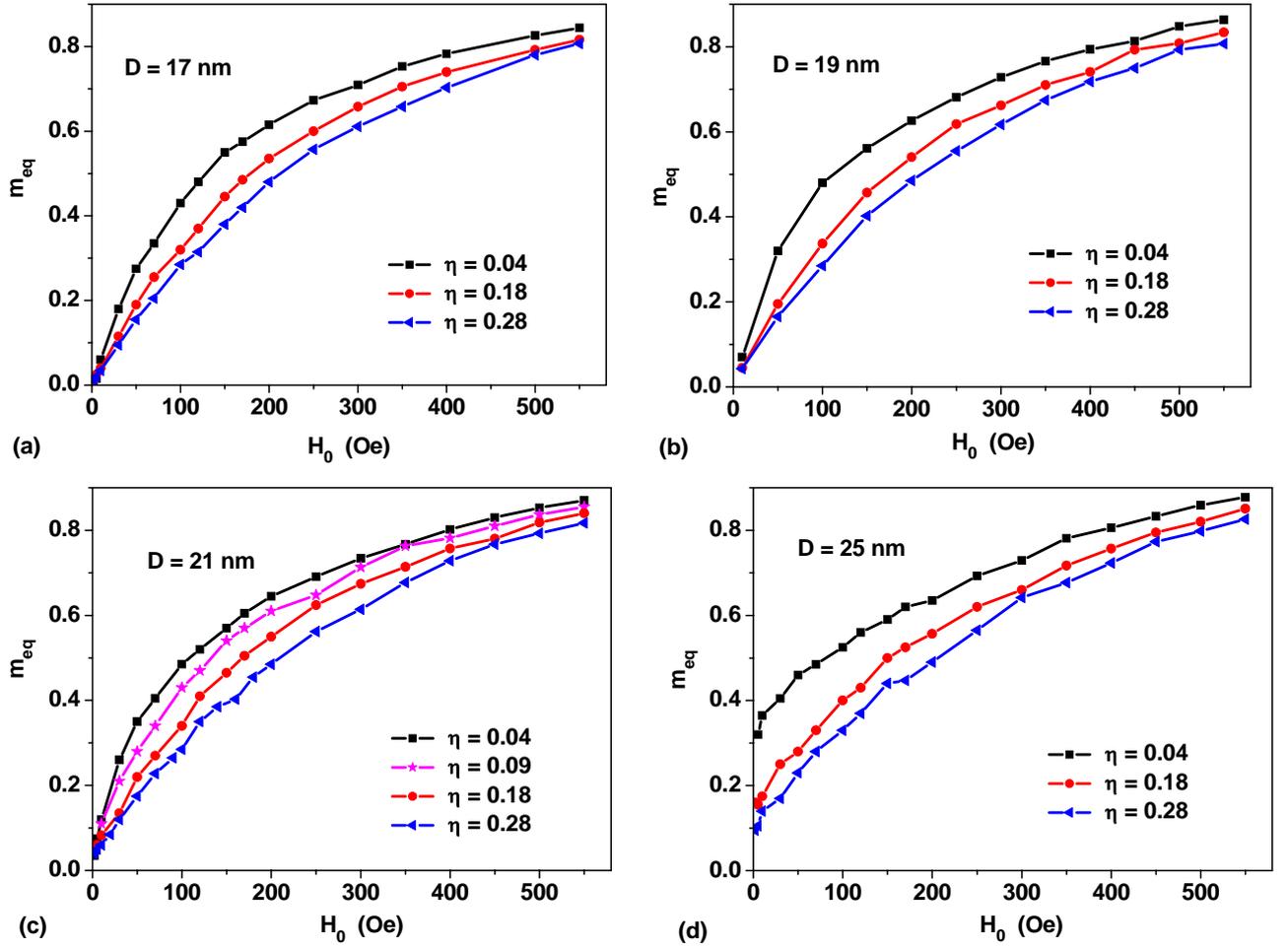

Fig. 3. Equilibrium reduced magnetization of dilute assemblies of clusters of magnetic nanoparticles of various diameters depending on the applied magnetic field at different cluster filling densities $\eta$. The assembly temperature is $T = 300°$ K, the magnetic anisotropy constant $K = 10^5$ erg/cm$^3$, the saturation magnetization $M_s = 350$ emu/cm$^3$, the number of particles in the clusters $N_p = 60$.

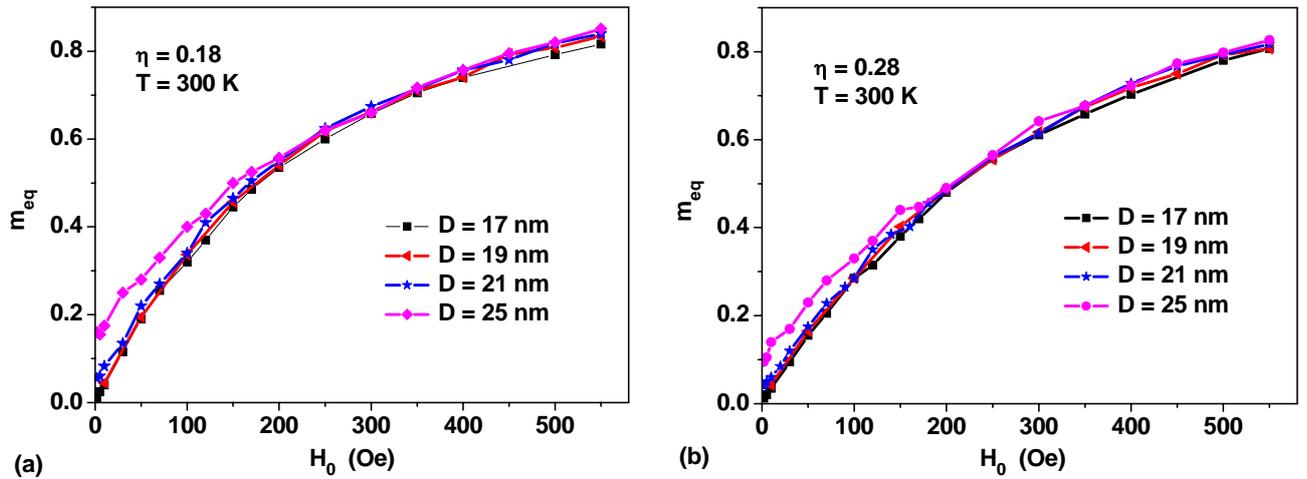

Fig. 4. Comparison of the reduced equilibrium magnetization for assemblies of nanoparticles of different diameters, but with the same cluster filling density $\eta$. Magnetic anisotropy constant $K = 10^5$ erg/cm$^3$, saturation magnetization $M_s = 350$ emu/cm$^3$.

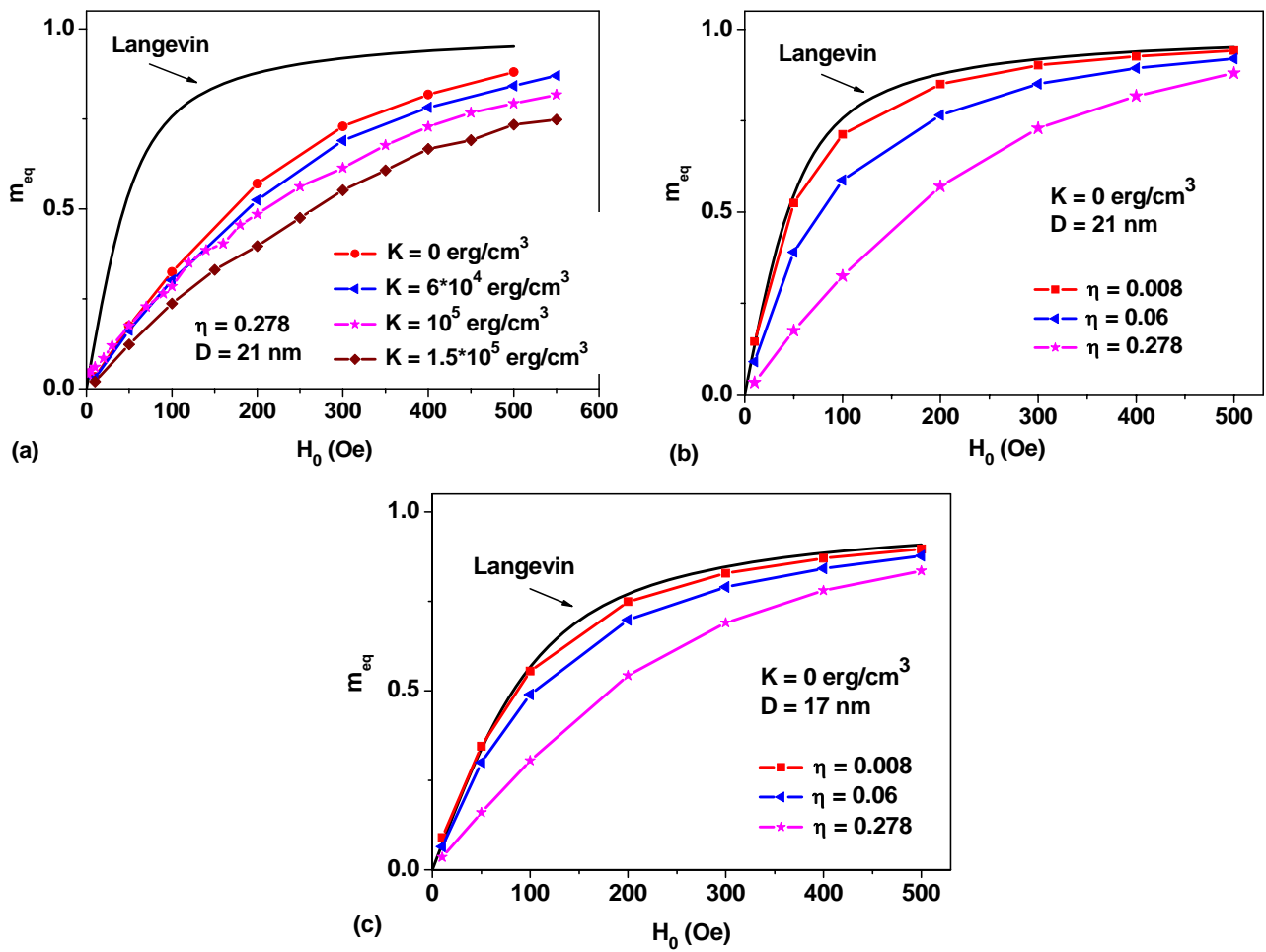

Fig. 5. Dependence of the equilibrium reduced magnetization: a) on the value of the magnetic anisotropy constant for assemblies with fixed cluster filling density $\eta = 0.278$; b), c) on the cluster filling density for an assembly of nanoparticles of various diameters with negligibly small anisotropy constant, $K = 0$.

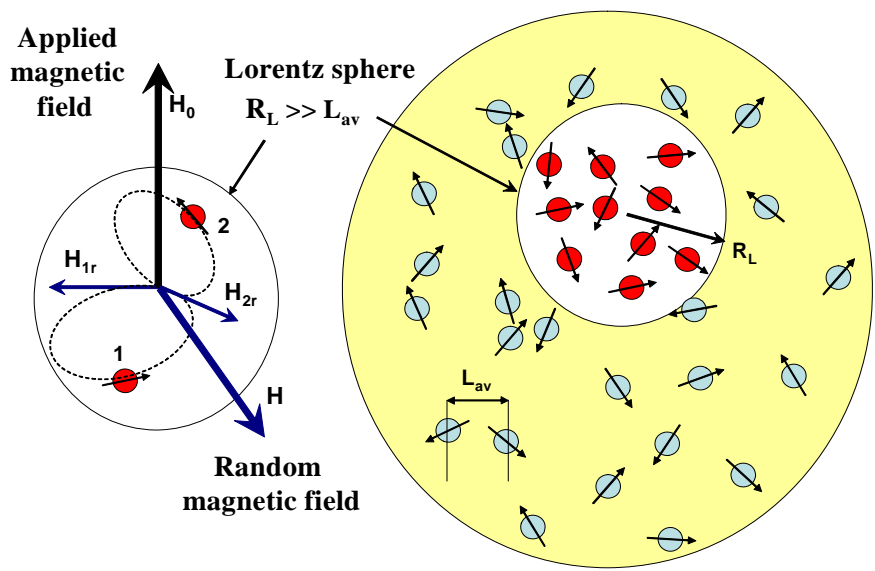

Fig. 6. Lorentz sphere around a reference nanoparticle in a large assembly of superparamagnetic nanoparticles.

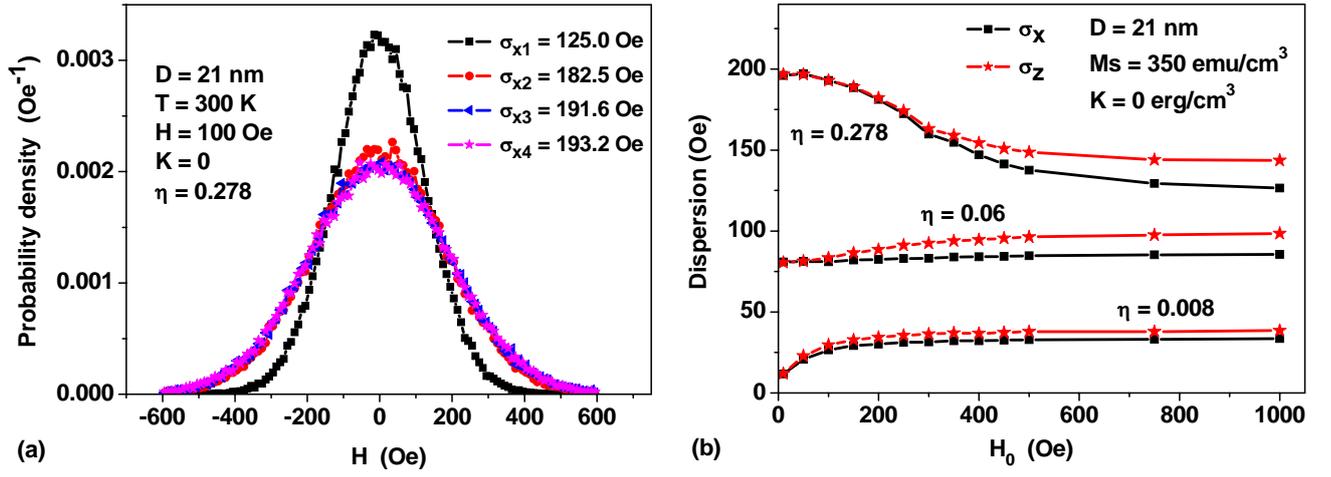

Fig. 7. a) Evolution of the empirical probability density $P^{(i)}(H_x)$ of the $H_x$ component of random magnetic field acting on a test nanoparticle located in the center of the Lorentz sphere for successive iterations $i = 1 - 4$; b) Limiting empirical standard deviations of the probability densities of the $H_x$ and $H_z$ components of random magnetic field for assemblies of nanoparticles with different filling densities $\eta$ as the functions of applied magnetic field.

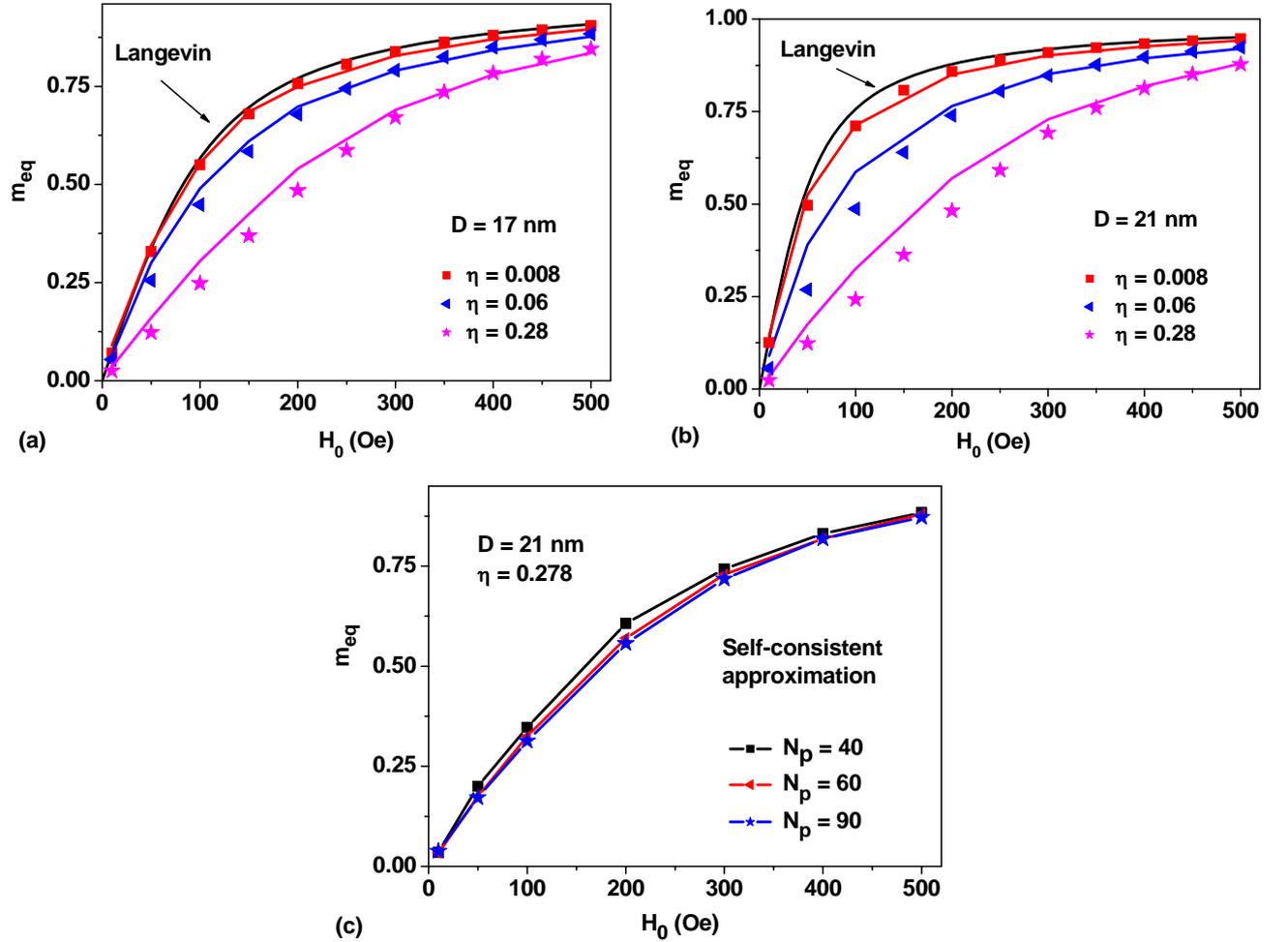

Fig. 8. Comparison of the equilibrium magnetizations of an assembly of nanoparticles with $K = 0$, $M_s = 350$ emu/cm$^3$ calculated by solving the stochastic LL equation (solid lines), and obtained in the self-consistent field approximation (dots) for particles of various diameters: a) $D = 17$ nm; b) $D = 21$ nm; c) comparison of the equilibrium assembly magnetizations obtained in the self-consistent approximation for different numbers $N_p$ of nanoparticles in the Lorentz sphere.

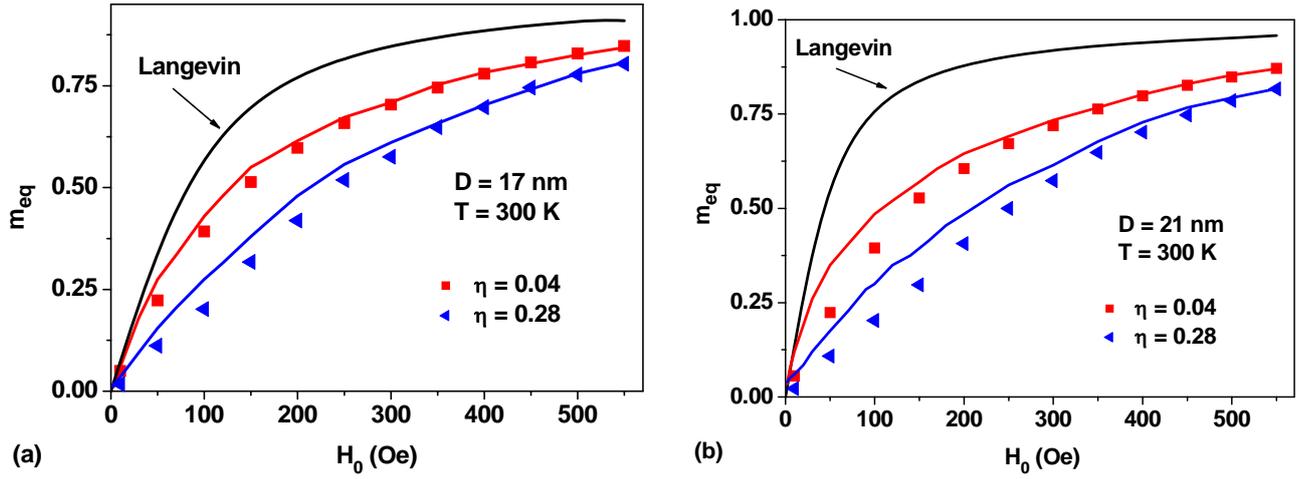

Fig. 9. Comparison of the equilibrium magnetization of assembly of random clusters of nanoparticles calculated by solving the stochastic LL equation (solid lines) with the corresponding results obtained in the self-consistent field approximation (dots) for particles of different diameters: a) $D$ = 17 nm, b) $D$ = 21 nm. Magnetic anisotropy constant $K = 10^5$ erg/cm$^3$, saturation magnetization $M_s$ = 350 emu/cm$^3$.

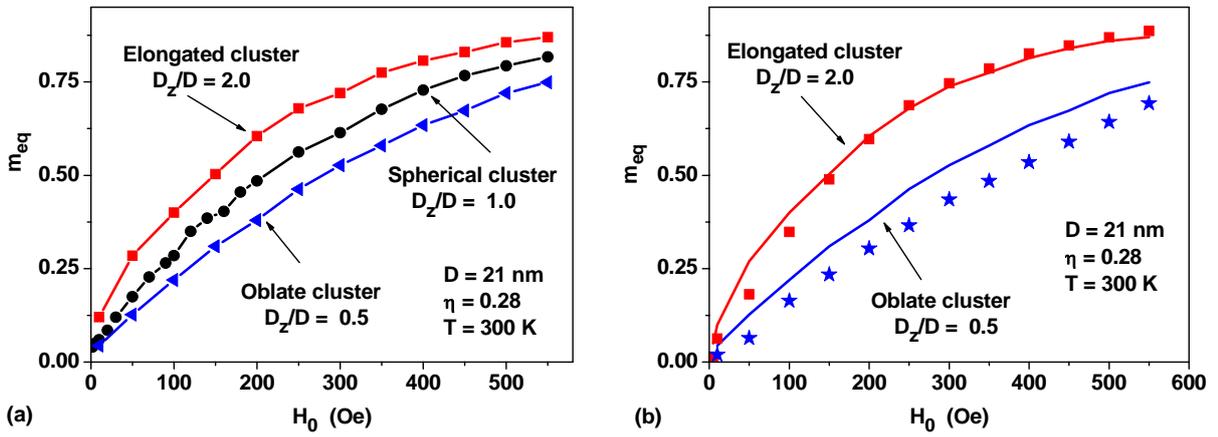

Fig. 10. a) The equilibrium magnetization of dilute assemblies of spheroidal clusters with different aspect ratios $D_z/D$ calculated by solving the stochastic LL equation; b) comparison of the results obtained by solving the stochastic LL equation (solid curves) with the corresponding calculations in the self-consistent field approximation (dots) for spheroidal clusters with different aspect ratios.